\documentclass[]{llncs}
\usepackage[utf8x]{inputenc}
\usepackage[numbers]{natbib}
\usepackage{graphicx}
\usepackage{listings}
\usepackage{hyperref}
\usepackage{amsmath}
\usepackage{multicol}
\usepackage{color}
\usepackage{tikz}
\usepackage{pgfplots}
\usepackage{tabularx}
\usepackage{booktabs}

\hypersetup{%
	colorlinks,
	linkcolor = black, 
	pdfpagemode = UseNone
}

\include{./images/q-grams.tex}

\title{Approximate Two-Party Privacy-Preserving String Matching with Linear
Complexity}

\author{Martin Beck\inst{1} \and
        Florian Kerschbaum\inst{2}}
		
\institute{Technische Universit\"at Dresden\\ Institute of Systems
		Architecture\\ Dresden, Germany\\ \email{martin.beck1@tu-dresden.de} \and
		SAP Research\\ Karlsruhe, Germany \\ \email{florian.kerschbaum@sap.com}}

\date{\today}
\begin{document}

\maketitle

\begin{abstract}
Consider two parties who want to compare their strings, e.g., genomes, but do
not want to reveal them to each other. We present a system for
privacy-preserving matching of strings, which differs from existing systems by
providing a deterministic approximation instead of an exact distance. It is
efficient (linear complexity), non-interactive and does not involve a third
party which makes it particularly suitable for cloud computing. We extend our
protocol, such that it only reveals whether there is a match and not the exact
distance. Further an implementation of the system is evaluated and compared
against current privacy-preserving string matching algorithms.
\end{abstract}


\section{Introduction}
\label{sec:intro}

As technology for sequencing the human genome is developing at a fast pace and
the number of sequenced genomes is rapidly growing, the need to process this
highly personalized information in a privacy preserving way also increases.
Several studies demonstrate how genomes can be linked to surnames
\cite{Gitschier2009a} or even reveal the full identity \cite{Gymrek2013,
Lunshof2008} of the individual. Many algorithms were presented which should
protect the genomic information while it is being processed across untrusted
parties.

These protocols however are either interactive, match only exact strings or
require a third party to be involved.
Our protocol is non-interactive, implements approximate string matching and does not require
any third party. 
Our protocol is efficient and has linear complexity in computation and
communication. It also has better resistance to an iterated differential attack proposed by
Goodrich \cite{Goodrich2009}, that exploits the information gained by knowing
the exact string distance (as proposed in other protocols), since it only reveals 
whether there is a match.
\vspace{1cm}
\subsection*{Our contributions:}
\begin{itemize}
  \item A new efficient privacy-preserving, non-interactive, two-party string
  matching protocol
  \item An analysis of our scheme in a genome matching setting using full
  mitochondrial DNA sequences
  \item We can privacy-preservingly, approximately compare real-world genomes 
  in under $5$ minutes on commodity hardware.
\end{itemize}

The remainder of this paper is structured as follows. Section
\ref{sec:relatedWork} gives an overview over basic concepts and related work. In 
Section~\ref{sec:protocolDesign} the design will be presented and followed by 
Section~\ref{sec:analysis}, which gives a security analysis of our system. Section
\ref{sec:evaluation} describes some implementation details and results in
comparison to related systems. Section \ref{sec:conclusion} concludes this work
and points out further research directions.

\section{Related Work}
\label{sec:relatedWork}

Research into string matching algorithms is defined by a long list of proposed
algorithms over many years and for many different problems. String matching
itself is closely related to the distance between strings, which can be measured
by a large variety of means, ranging from generic and simple solutions like the
Hamming distance \cite{Hamming1950} to more powerful algorithms
like Smith-Waterman \cite{Smith1981} solving local sequence alignment problems.
A survey about current developments can be found in \cite{Li2010}.

\subsection{Approximate String Matching}
\label{subsec:approxStringMatching}

As several tasks, for example checking whether a user profile is within a remote
database, do not require the exact distance between two strings, data items or
other entities, the notion of approximate matching was introduced to define
levels of similarity, which in the most extreme way only output a single bit of
information: if the input strings are similar or not. Due to these properties
this class is called approximate string matching algorithms, which is not to be
confused with the approximate string matching of \cite{Hall1980}, where the term
``approximate'' referred to the property of two strings being close in distance.

\subsection{Privacy-Preserving String Matching}
\label{subsec:ppsm}

Two of the applications for string comparison algorithms which are often
used for motivation are calculating the distance of genome or protein sequences
in bioinformatics and checking if a person is present in a remote database. As
these topics by design deal with very personal information, which must not be
given to third parties, the necessity to build privacy-preserving matching
algorithms arose. As these were not sufficient to protect privacy due to
information leakage given by the exact distance results, just obtained in a
privacy preserving manner, combinations of the above mentioned approximation and
the privacy-preserving computational steps were developed. A survey of recently
published algorithms together with benchmark results can be found in
\cite{Bachteler2010}.

One of the more recent protocols introduced by Schnell et al.
\cite{Schnell2009} uses Bloom filters to represent strings and transforms the
notion of distances between strings into distances between similar Bloom filters. We
will also use Bloom filters as set representation for our genomic strings and
build the matching protocol upon them. 
However, we use a two-party technique for comparing the Bloom filters and
therefore do not need a trusted third party for comparing the strings.
Furthermore, our protocol can be size-hiding, by choosing appropriate
Bloom filter sizes, that are not proportional to the string length.

Alternatively, techniques from private set intersection (PSI) \cite{Huang2012}
could be used.
However, revealing the content of the intersection is not appropriate for a
privacy preserving protocol. Based on these security concerns, protocols for
private set intersection cardinality (PSI-CA) were developed \cite{Cristofaro}.
Yet, these solutions still reveal the intersection cardinality, whereas we only
reveal whether there is a match.

Privacy-preserving protocols designed for approximate string comparisons can
also be found in literature \cite{Troncoso-Pastoriza2007,Jha2008}, but rely on
interactive techniques like oblivious transfers or secure computation. 
This excludes these protocols from off-line execution, e.g., in the cloud.
Further \cite{Baldi:2011:CGE:2046707.2046785} presents a more
efficient solution, but which only matches exact strings, whereas we compare
approximate strings.

\section{Protocol Design}
\label{sec:protocolDesign}

Let the client (Alice) have a string, e.g.~a genome, and the server have a string.
After the execution of our protocol Alice will have learned whether the two strings
are approximately close, but not Bob's string nor the approximate distance to Bob's 
string.  Bob will learn nothing.

First the transformation into grams of variable length and their representation
through a Bloom filter is specified, upon which the generic string matching
algorithm is given. Following this generic matching algorithm a
privacy-preserving version is constructed and then enhanced to only reveal whether 
there is a match.

\subsection{Bloom Filter Representation}
\label{sec:bloomFilter}

A Bloom filter is a data structure fixed in size to which element
representations can be added and on which member tests can be performed. Checking for an
element is probabilistic due to the design of the filter.

Let $b$ be an array of bits of length $n$ and $b[i]$ the $i$th value within the
array with $i \in [1,n]$. Further let $h_1() \dots h_k()$ be $k$ hash
functions, with uniformly distributed output in $[1,n]$. For initialization
set $\forall i \in [1,n]: b[i] = 0$.

To add an element $e$ to the filter, all $k$ hash functions are evaluated on $e$
and the results are taken as indexes for $b$ to set these positions to one. Set
$\forall j \in [1,k]:  b[h_j(e)] = 1$.

A member test for element $e'$ is performed by also evaluating all $k$ hash
functions and checking the referenced positions in $b$. If at least one of the
positions $b[h_j(e')]$ is set to zero, the element has not been added to the
Bloom filter before. If all bits are set to one, however, one cannot be sure if the exact
element was inserted, or one or more different elements had these positions set
to one.

Using these operations a set is represented by adding all set elements
to the filter. Depending on the filter parameters, the probability that a
false-positive member test occurs, i.e. that an element is falsely identified as
being added to the filter before, is given by:

\begin{align}
p &= \left(1-\left(1-\frac{1}{n}\right)^{kl}\right)^k \label{eq:prob}
\end{align}

Where $\left(1-\frac{1}{n}\right)^{kl}$ is the probability that a single bit is
still zero after $l$ elements were added to the filter of length $n$ using $k$
hash functions. To calculate the required length of a Bloom filter $n$ given the
false-positive rate and the number of elements to be inserted $l$, the equation
\eqref{eq:prob} can be transposed to:

\begin{align}
n &= \frac{-1}{\left(1-p^{1/k}\right)^{1/(k*l)} -1} \label{eq:bloomLen}
\end{align}

\subsection{String Matching Using Bloom Filters}
\label{sec:stringMatching}

A typical string comparison algorithm is the Levenshtein distance
\cite{Levenshtein1966}, which is often also referred to as edit distance and
describes the minimum number of insertions, deletions and substitutions needed to transform
one string $S_1$ into another $S_2$. The result is a distance measure $d$, which can
easily be converted into a similarity score $s$ between zero and one by: $s =
1-\frac{d}{d_{max}}$

$d_{max}$, i.e. the maximum distance between two strings,
equals the length of the longer string and can thus be replaced by $d_{max} =
\text{ max}(|S_1|,|S_2|)$ regarding the Levenshtein distance.

$$
s = 1-\frac{d}{\text{ max}(|S_1|,|S_2|)}
$$

As a Bloom filter is a set representation, the input strings first need to be
converted into sets. This has to be done in a way, that a distance measure can
later be formulated upon the constructed set which, loosely spoken, correlates
with the Levenshtein distance measure.

The sets are build from $q$-grams, which are substrings of length $q$ from input
string $S$. Let $n=|S|$ be the number of characters in
$S$ and $s_i$ the $q$-gram starting at position $i$ with $i \in [1,n-q+1]$. As a
result $n-q+1$ $q$-grams are generated out of $S$ using a sliding window for
all possible $i$. If this set would be used to represent a string and measure
similarity upon, the positional information of the substrings would not be
included, which is important to build the similarity measure. To keep this
information positional $q$-grams are used, which are pairs $(i,s_i)$ with $i$ being the
position in $S$ and $s_i$ the actual $q$-gram starting at that position.

Further as characters at the beginning and at the end of $S$ are
underrepresented over all $q$-grams, the input string $S$ is extended by $q-1$
identical symbols, which are not part of the alphabet of $S$ at the beginning
and end of $S$. Gravano et al. \cite{Gravano2001} introduced this definition of
positional $q$-grams on extended strings. Without the extension we would only see the first
character in the first $q$-gram, whereas in the middle of the string each
character is found within $q$ $q$-grams.

These positional grams are not used directly, but a technique called VGRAM
\cite{Li2007} is employed to generate grams with variable lengths within
a previously defined range $[q_{min},q_{max}]$. To choose which length to select
at what position, a gram dictionary is build prior to running the string
comparison algorithm. As source to build this dictionary, the Human
Mitochondrial Genome Database \cite{Ingman2006} is used. Afterwards the
generated dictionary is published and available to all participants described in
this string matching algorithm.

Both parties use the VGRAM algorithm to build a set of variable length grams
based on the published dictionary, following the description in \cite{Li2007}.
The number of variable length grams generated for a string $S$ is depicted by
$n_v$.

%


As these sets cannot be used directly to compare strings in a
privacy-preserving way, which would reveal the original data, we represent
them using Bloom filters. A single Bloom filter is used for all grams generated
by a single string $S$. Papapetrou et al. \cite{Papapetrou2010} conclude, that
the optimal number of hash functions to do cardinality estimation using Bloom
filters is $1$. Based on this we fix $k = 1$ and only use a single hash function
to build and query Bloom filters throughout the rest of the paper. The length
$l$ of the Bloom filter and the used hash function $h()$ is also fixed and set
to be equal across all participants.

Determining an appropriate value for $l$ can be done using the formula
\eqref{eq:bloomLen} with the simplification $k = 1$ as introduced above. This
results in $l$ being calculated as:

\begin{align}
l &= \frac{-1}{\left(1-p\right)^{n_v^{-1}} -1} \label{eq:bloomLenK}
\end{align}

Under these constraints, that $k$, $l$ and $h()$ are identical, set union $\cup$
and intersection $\cap$ can also be performed upon Bloom filters $B_1,B_2$ by
applying the binary OR or AND operator.

Li et al. \cite{Li2007} describe the effect of a single edit operation on the
set of variable length grams and propose an algorithm to calculate the maximum
number of affected grams by applying a number of edit operations upon an initial
string. Based on this number a lower bound on the number of
common grams for two strings which are within a certain edit distance can be
calculated.

As this lower bound calculation on the set intersection cardinality uses the
input sequences to generate a baseline for the lower bound, it cannot be
applied in our privacy-preserving scheme directly. To be independent of
such a baseline, we do not use the absolute set intersection cardinality
directly, but the difference between the union cardinality and the intersection
cardinality. This results in an approximate distance measure, which follows the
explanation for the upper bound on the Hamming distance between bit vectors in
\cite{Li2007}. Our distance measure equals the Hamming distance between the
Bloom filter bit vectors.

$$
d = |B_1 \cup B_2| - |B_1 \cap B_2|
$$

Thus having $d = 0$ equals to having identical strings, as the set union and
intersection sizes are identical.

\subsection{Encrypting the Bloom Filter}
\label{sec:encBloom}

We constructed string representations using Bloom filters in Section~\ref{sec:stringMatching}
and used them to define an appropriate distance
measure. However the Bloom filters themselves cannot be exchanged between the
participants directly, as they can be used to possibly reconstruct the original
strings by guessing substrings and checking if they were added to the filter.
For preserving privacy of the filter content, an additively
homomorphic cryptosystem is used.

A homomorphic cryptosystem uses at least one homomorphic property to evaluate an
operation $\oplus$ on the ciphertext, which translates into applying the
equivalent operation $+$ on the plaintext. We will use the additively
homomorphic system introduced by Naccache and Stern~\cite{Naccache1998}, which
is also probabilistic. Alice generates a key pair and shares the public key with
Bob. Let $E(x, r)$ denote the encryption of a value $x$ using a fresh random
value $r$ for each encryption, this additively homomorphic system has the
following properties:

\begin{eqnarray*}
E(x,r)E(y,s) &=& E(x + y,rs)\\
E(x,r)^y &=& E(xy, r^y)
\end{eqnarray*}

Further let $E(x,r)^{-1}$ denote the calculation of the multiplicative inverse
upon $E(x,r)$, found through executing the extended euclidean algorithm, which
is by the homomorphism definition the encryption of the additively inverse
plaintext. This results in $E(x,r)^{-1} = E(-x, r)$ and can be used to
calculate a difference between two encrypted values.

To multiply an encrypted plaintext with a negative factor $-z$, first the
multiplicative inverse of the encrypted value is calculated and then multiplied
using the positive factor. $E(x,r)^{-z} = E(x, r)^{-1\cdot z} = E(-x,r)^{z} =
E(-xz,r^z)$. To increase readability, $E(x) = E(x,r)$ is used, which also
always uses a fresh $r$.

An encryption of a Bloom filter $B$ with length $l$ is constructed by encrypting
every bit in $B$ separately, storing the resulting $l$ values in a new array $C$
with equal length.

$$
\forall i \in [1,l] \text{, fresh } r: \quad C[i] = E(B[i], r)
$$

This encryption is not to be confused with ``encrypted Bloom filter'', which
were presented in \cite{Bellovin2004}. Encryption of the Bloom filter is only
performed by Alice, who wants to compare a string privately to one that Bob
holds. Bob also slices his string down into variable length grams, which are
then added to a new Bloom filter using the previously agreed upon parameters
$k$, $l$ and $h()$.

Alice sends the encryption of her Bloom filter to Bob, together with her public
key. Recall that the Bloom filters just contained zeros and ones. So calculating
the cardinality of a filter, denoted by $|B|$, cannot just be done by counting
all bits set to one, but also by calculating the sum over all values $|B| =
\sum_{i=1}^{l} B[i]$.

Bob can use this property to calculate the encrypted sum over all values in the
encrypted Bloom filter and thus the encrypted cardinality.
$$
E(|B|,r) = E(\sum_{i=1}^{l} B[i],r) = \prod_{i=1}^{l} C[i]
$$

However as Bob is not interested in the encrypted cardinality of Alice's filter
$|B_A|$, he only adds up values at those positions, that are set to one on his
own Bloom filter $B_B$. This is equivalent to building the intersection using
binary AND and calculating the resulting cardinality. 

$$
E(|B_A \cap B_B|) = \prod_{i, B_B[i]=1} C[i]
$$

Further Alice encrypts the cardinality of her Bloom filter $E(|B_A|)$ and sends
it to Bob, who also encrypts the cardinality of his own Bloom filter $E(|B_B|)$.
Using these values, the union cardinality is calculated as follows:

\begin{eqnarray*}
|B_A \cup B_B| &=&  |B_A \cap B_B| + (|B_A| - |B_A \cap B_B|) + (|B_B| - |B_A
\cap B_B|)\\
 &=& |B_A| + |B_B| - |B_A \cap B_B|
\end{eqnarray*}

\begin{eqnarray}
E(|B_A \cup B_B|) &=& E(|B_A|) \cdot E(|B_B|) \cdot E(|B_A \cap B_B|)^{-1}
\end{eqnarray}

This way the encrypted distance $E(d)$ for the measure presented in Section~\ref{sec:stringMatching} is calculated as such:

\begin{eqnarray}
E(d)	&=& E(|B_A \cup B_B|) \cdot E(|B_A \cap B_B|)^{-1}  \\
		&=& E(|B_A|) \cdot E(|B_B|) \cdot E(|B_A \cap B_B|)^{-2}
\end{eqnarray}

\subsection{Privacy-Preserving Similarity}
\label{sec:pps}

The calculated approximate distance value $E(d)$ between both compared strings
$S_A$ and $S_B$ in Section~\ref{sec:encBloom}, could be returned to Alice for
decryption, for her to learn the actual computed value. This would however
result in increased sensitivity to the Mastermind attack described in
\cite{Goodrich2009}. To circumvent this attack, we restrict the information
Alice gains from executing this protocol. Instead of learning the exact result
of the comparison, the result is manipulated to give Alice only the information
whether the distance is smaller than a previously defined threshold $t_{max}$.

Recall that the calculated distance value equals the Hamming distance between
the Bloom filters and that \cite{Li2007} describes how to calculate an upper
bound for the Hamming distance in equation $(4)$. The calculation however
involves the number of affected grams for both input strings $S_A$ and $S_B$.
As $S_B$ is not available to Alice, she uses the revised Cambridge Reference
Sequence (rCRS) \cite{Anderson1981} $S_{rCRS}$ as a reference to replace $S_B$
in the calculation of the upper Hamming distance bound. This replacement is a
good approximation for small edit distances. Following \cite{Li2007} the
upper bound for a maximum edit distance $e_{max}$ is calculated as $t_{max} =
NAG(S_A, e_{max}) + NAG(S_{rCRS}, e_{max})$, where $NAG(S,e)$ describes the
maximum number of affected grams for $e$ edit operations on string $S$. Further,
as $NAG(S_{rCRS}, e_{max})$ is very close to $NAG(S_B, e_{max})$ and thus used
as an replacement. $NAG(S_A, e_{max})$  can also be replaced for the same
reason. This has the effect, that the chosen $t_{max}$ does not depend
on the input sequence $S_A$.

Due to the probabilistic nature of the Bloom filter, elements are mapped to
the same positions with a probability $p$ as described in 
Section~\ref{sec:bloomFilter}. As the Bloom filter cardinality $|B_A|$ is therefore on
average smaller than the number of variable grams for $S_A$ by a factor $p$, the
upper bound is corrected to an approximated upper bound.

\begin{eqnarray}
t_{max} &=& 2 \cdot NAG(S_{rCRS}, e_{max})) \cdot (1-p)
\end{eqnarray}

%
%

%
%

\noindent
The protocol for calculating the return values for Alice by Bob
is as follows:
\begin{itemize}
  \item Calculate encrypted inverse thresholds $\forall t_i \in
  \left[0,t_{max}\right]: E(-t_i) = E(t_i)^{-1}$
  \item Calculate encrypted threshold differences for all inverse thresholds\\ $
  E(D_i) = E(d-t_i) = E(d) \cdot E(-t_i)$
  \item Multiply all differences with random values. $E(rD_i) = E(D_i)^r$ for
  fresh $r$ drawn uniformly from the plaintext space of the used cryptographic
  system.
\end{itemize}

After the first two steps Bob has $t_{max}+1$ values, expressing
the differences between the incremented thresholds and the actual distance. If the
calculated distance $d$ is within the defined threshold range
$\left[0,t_{max}\right]$, then there is one single element, which is the
encryption of zero due to equal threshold and distance values.

Performing the last step randomizes all values through multiplication with a
random number, except the one encrypting a zero. All these $t_{max}+1$
encrypted values are then shuffled randomly and sent to Alice, who decrypts
and checks them against zero. In case a zero is found, she learns that the Bloom
filter intersection cardinality was within the specified threshold and thus the
compared strings have an edit distance equal or less than the specified maximum
edit distance $e_{max}$ used to calculate $t_{max}$ in 
Section~\ref{sec:stringMatching}.

\section{Security Analysis}
\label{sec:analysis}

Our protocol is secure under the semi-honest, also called honest-but-curious
model and under the assumption the integrated crypto system builds upon. In our
case this is based on the higher residuosity problem used in the Naccache-Stern
cryptosystem. Several other additive homomorphic cryptosystems like Paillier~\cite{Stern1999} can easily
be used instead of the currently employed system, bringing possibly another
assumption like one based on the decisional composite residuosity problem as
basis.

The encryption of the used cryptosystem must however be probabilistic, such that
similar plaintexts are mapped to different ciphertexts at random. This is true
for our employed Naccache-Stern system and the above mentioned Paillier
cryptosystem.
This property is also called semantic security and corresponds to indistinguishability
under chosen plaintext attack (IND-CPA).

In the first part of our protocol, Alice translates her input string into
variable length grams, generates a Bloom filter representation and encrypts it
using a public key cryptosystem. As she is not using any
information from Bob, she cannot gain any insight into Bob's input.

The second part involves Bob working on the encrypted Bloom filter from Alice
and her encrypted Bloom filter cardinality. As all values are encrypted using an
asymmetric, probabilistic cryptosystem, for which only Alice has the private
decryption key, Bob cannot decide if an encrypted value represents a zero, a
one or any other value, which directly follows the security analysis of the
underlying hardness assumption. The number of elements received does not depend
on Alice input, as only public information is used to infer the Bloom filter
length, as introduced in Section~\ref{sec:pps}. Further Bob sums up elements
from Alice's encrypted Bloom filter, based on his Bloom filter. The result is
then subtracted several times from different threshold values and multiplied
with random numbers, chosen uniformly from within the domain of plaintexts of
the underlying cryptosystem. All results are shuffled at random and transmitted
back to Alice. Bob gained no information in this phase about Alice's input.

As a last step Alice decrypts all results received from Bob and checks if they
contain a zero. If a zero is found, she learns that the Hamming distance between
the Bloom filters was below a predefined threshold $t_{max}$. There can be at most 
one zero. If no zero was
found, the threshold was lower than the Hamming distance. From the decrypted
non-zero results, she cannot learn anything, as these numbers are uniformly
distributed due to the multiplication with uniform random numbers drawn from the plaintext
domain modulo the the plaintext domain modulus. The index of the zero element,
if there was one, gives no information to Alice, as the return values were
randomly shuffled by Bob. The number of returned elements also holds no further
information, as there are always $t_{max}+1$ results.

The only information Alice learns about the input of Bob is, if the threshold
was above the Bloom filter Hamming distance or not.

\section{Evaluation}
\label{sec:evaluation}

For the experiments a Linux Laptop with an Intel Core$2$ Duo T$9600$
running at $2.8$ GHz was used. The code is written in Java, using the Bouncy
Castle library\footnote{http://www.bouncycastle.org/}. The first tests evaluate
the relation between the Levenshtein distance and the measure introduced in
section \ref{sec:stringMatching}. Further the runtime performance of
the algorithm is evaluated for string lengths, which were also used for
comparing other privacy-preserving string matching protocols. All code
implementing the techniques in this paper and producing the test results can be
found under \url{http://dud.inf.tu-dresden.de/~beck/bloomEncryption.tar.bz2}.

\subsection{Distance Measure}
\label{sec:dm}

As the similarity metric is based on the Levenshtein distance as described in
\cite{Li2007}, we measure the relation between the edit distance and the
Bloom filter Hamming distance as our proposed metric. The parameters $q_{min} =
2$ and $q_{max} = 40$ are used as \cite{Li2007} states that the variable length
gram algorithm can start with a low $q_{min}$ and a large $q_{max}$ to find
appropriate values for these parameters after pruning the built Trie.

To run the tests the Bloom filters are set up to use a single hash function, in
our case ``SHA$1$'' modulo the size of the filter. The used strings contain
roughly $n=16,569$ characters, which means that about the same number of
variable grams have to be inserted into each Bloom filter.

The probability that a false-positive test occurs after the $n$ elements are
added to the filter is set to $p=0.1$, which in turn generates a Bloom filter of
size $157261$ bits. We used $10000$ runs and for each $100$ applied a fixed
number of edit operations. The original string and the altered string are then
compared using our distance measure. The resulting value is the difference
between the union cardinality and the intersection cardinality of both Bloom
filters $B_1$ and $B_2$. This represents the total number of unique elements for
both parties, or the Hamming distance between both Bloom filters. Following
Lemma $1$ in \cite{Li2007} this directly correlates with the Levenshtein
distance between the strings.

\begin{figure}[h!]
  \label{fig:distances}
  \centering
    \includegraphics[width=0.6\textwidth]{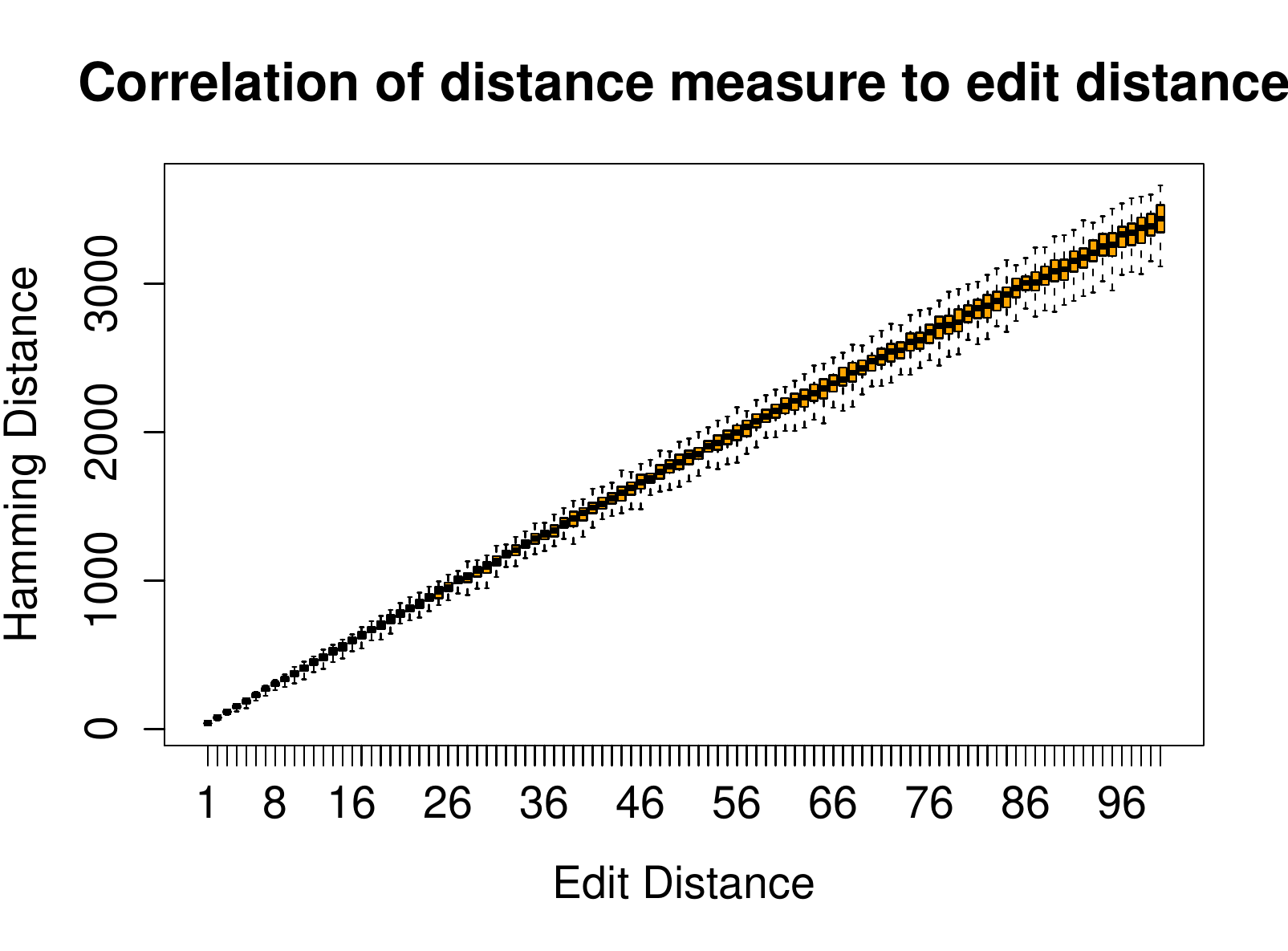}
  \caption{Approximation of the Levenshtein distance by our distance measure}
\end{figure} 

Figure \ref{fig:distances} shows a Boxplot for every Levenshtein distance and
the according $100$ runs tested with our approximate distance measure. As can
be seen from the figure, our distance value approximates small Levenshtein
distances very good, with a narrow range of possible values and a small
variance. The Pearson correlation between the Levenshtein distances and the
approximated distances is $c_p = 0.997$ for up to $100$ edit operations.



\subsection{Protocol Execution Time}
\label{sec:pet}
To evaluate the performance of our protocol, we ran $100$ runs for each
test. The parameters were set to $q_{min}=2$, $q_{max}=40$, $p=0.1$ and
an edit distances of up to $10$ operations.

The client runtime depends linearly on the length of the input sequence, where
the most time is spent on decrypting the results from the server and encrypting
the Bloom filter prior to transmission. We can see a pretty high variance on
client runtimes, growing linearly with longer sequences, due to the unknown
number of results which are needed to be decrypted until a zero is found. If the
distance between both compared sequences is not within the predefined range
given by the threshold, the client always needs to decrypt all results, as no
zero will be found within the returned values.

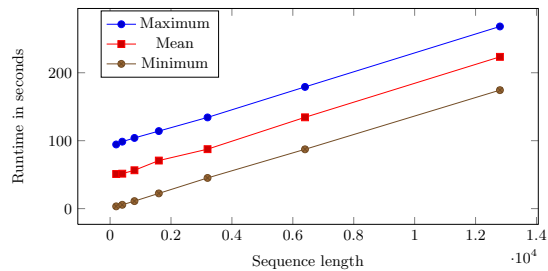
\begin{figure}[h!]
  \centering
  \label{fig:clientRuntime}
	\begin{tikzpicture}[scale=.65]
	\begin{axis}[xlabel={Sequence
	length},ylabel={Runtime in seconds},legend entries={Maximum,Mean,Minimum},
	legend style={at={(0.05,0.68)},anchor=south west}, height=6cm, width=11cm]
	\addplot coordinates {
		(200,94.42)
		(400,98.54)
		(800,104.03)
		(1600,114.13)
		(3200,134.24)
		(6400,179.14)
		(12800,268.12)
	};
	\addplot coordinates {
		(200,50.93)
		(400,51.45)
		(800,56.44)
		(1600,70.74)
		(3200,87.5)
		(6400,134.39)
		(12800,223.4)
	};
	\addplot coordinates {
		(200,3.33)
		(400,5.62)
		(800,11.03)
		(1600,22.41)
		(3200,45.16)
		(6400,87.29)
		(12800,174.38)
	};
	\end{axis}
	\end{tikzpicture}
  \caption{Client runtimes}
\end{figure} 

Server runtime depends linearly on the threshold value, whereas runtimes for
different sequence lengths are only increasing slightly. The measured values for
a constant threshold derived from a maximum edit distance of $10$ and a variable
sequence length range between $8.54$ seconds for sequences of length $200$ and
$9.01$ seconds for full mitochondrial DNA sequences.


The amount of data that needs to be transferred between Alice and Bob is shown
in table \ref{tbl:bandwidth} and grows linearly with the length of the Bloom
filter for the traffic from Alice to Bob and linearly with the size of the
threshold range for the traffic from Bob to Alice. For this test the threshold
$t_{max}$ is set to the maximum Hamming distance defined in 
Section~\ref{sec:pps} for a maximum edit distance of $10$.

\begin{table}[h!]
\centering
\scriptsize
\begin{tabular}{r r r}
\toprule
Sequence length & Client to server & Server to client \\
 \midrule
  $200$ & $296$ KB & $123$ KB \\
  $400$ & $590$ KB & $123$ KB \\
  $800$ & $1169$ KB & $123$ KB \\
 $1600$ & $2337$ KB & $123$ KB \\
 $3200$ & $4663$ KB & $123$ KB \\
 $6400$ & $9323$ KB & $123$ KB \\
$12800$ & $18636$ KB & $123$ KB \\
\bottomrule 
\end{tabular}
\caption{Bandwidth used for transmission}
\label{tbl:bandwidth} 
\end{table} 

Comparing these results to the ones given by Jha et al.~\cite{Jha2008} and Huang
et al.~\cite{Huang2011} in the evaluations of their state of the art protocols, we
achieve an increased performance starting with the smallest sequence lengths of
$200$ characters. Due to the lower linear complexity of our protocol,
comparisons of full mitochondrial DNA sequences can be performed more
efficiently. The referenced protocols have computational complexity of $O(n
\text{ log } n)$, $O(n^2)$ and $O(n*m)$ for input string lengths $n$ and $m$.

\section{Conclusion}
\label{sec:conclusion}

We presented a novel, non-interactive approach for a privacy-preserving
approximate string matching protocol, that achieves superior performance 
for real-world sized genomes.
An attacker will not even learn the exact distances or approximations, but only 
whether two compared strings are within a predefined distance range.

Due to the computation having linear complexity in the used sequence length
and the communication having linear complexity in the range of allowed
distances, respectively in the Bloom filter length, this protocol is very
practical and was tested for full mitochondrial sequences with $16500$
characters in length and a maximum edit distance of $10$, which took about $286$
seconds on the mentioned hardware to complete.

Further enhancements 
could go into using our protocol for database searches.

\bibliographystyle{plainnat2}
\bibliography{./library}

\begin{thebibliography}{25}
\providecommand{\natexlab}[1]{#1}
\providecommand{\url}[1]{\texttt{#1}}
\expandafter\ifx\csname urlstyle\endcsname\relax
  \providecommand{\doi}[1]{doi: #1}\else
  \providecommand{\doi}{doi: \begingroup \urlstyle{rm}\Url}\fi

\bibitem[Anderson et~al.(1981)Anderson, Bankier, Barrell, de~Bruijn, Coulson,
  Drouin, Eperon, Nierlich, Roe, Sanger, Schreier, Smith, Staden, and
  Young]{Anderson1981}
S.~Anderson, A.~T. Bankier, B.~G. Barrell, M.~H.~L. de~Bruijn, A.~R. Coulson,
  J.~Drouin, I.~C. Eperon, D.~P. Nierlich, B.~A. Roe, F.~Sanger, P.~H.
  Schreier, A.~J.~H. Smith, R.~Staden, and I.~G. Young.
\newblock {Sequence and organization of the human mitochondrial genome}.
\newblock \emph{Nature}, 290\penalty0 (5806):\penalty0 457--465, April 1981.
\newblock ISSN 0028-0836.

\bibitem[Bachteler and Schnell(2010)]{Bachteler2010}
Tobias Bachteler and Rainer Schnell.
\newblock {An empirical comparison of approaches to approximate string matching
  in private record linkage}.
\newblock \emph{Proceedings of Statistics Canada}, 2010.

\bibitem[Baldi et~al.(2011)Baldi, Baronio, {De Cristofaro}, Gasti, and
  Tsudik]{Baldi:2011:CGE:2046707.2046785}
Pierre Baldi, Roberta Baronio, Emiliano {De Cristofaro}, Paolo Gasti, and Gene
  Tsudik.
\newblock {Countering GATTACA: efficient and secure testing of fully-sequenced
  human genomes}.
\newblock In \emph{Proceedings of the 18th ACM conference on Computer and
  communications security}, CCS '11, pages 691--702, New York, NY, USA, 2011.
  ACM.
\newblock ISBN 978-1-4503-0948-6.

\bibitem[Bellovin et~al.(2004)Bellovin, Bellovin, and Cheswick]{Bellovin2004}
Steven~M. Bellovin, Steven~M Bellovin, and William~R Cheswick.
\newblock {Privacy-Enhanced Searches Using Encrypted Bloom Filters}, 2004.

\bibitem[Cristofaro et~al.(2011)Cristofaro, Gasti, and Tsudik]{Cristofaro}
Emiliano~De Cristofaro, Paolo Gasti, and Gene Tsudik.
\newblock {Fast and Private Computation of Cardinality of Set Intersection and
  Union}.
\newblock \emph{Cryptology ePrint Archive, Report 2011/141}, pages 1--19, 2011.

\bibitem[Gitschier(2009)]{Gitschier2009a}
Jane Gitschier.
\newblock {Inferential genotyping of Y chromosomes in Latter-Day Saints
  founders and comparison to Utah samples in the HapMap project.}
\newblock \emph{American journal of human genetics}, 84\penalty0 (2):\penalty0
  251--8, February 2009.
\newblock ISSN 1537-6605.

\bibitem[Goodrich(2009)]{Goodrich2009}
Michael~T. Goodrich.
\newblock {The Mastermind Attack on Genomic Data}.
\newblock In \emph{2009 30th IEEE Symposium on Security and Privacy}, pages
  204--218. IEEE, May 2009.
\newblock ISBN 978-0-7695-3633-0.

\bibitem[Gravano et~al.(2001)Gravano, Ipeirotis, Jagadish, Koudas,
  Muthukrishnan, and Srivastava]{Gravano2001}
Luis Gravano, Panagiotis~G. Ipeirotis, Hosagrahar~Visvesvaraya Jagadish, Nick
  Koudas, Shanmugauelayut Muthukrishnan, and Divesh Srivastava.
\newblock {Approximate String Joins in a Database (Almost) for Free}.
\newblock \emph{Proceedings of the 27th International Conference on Very Large
  Data Bases}, pages 491--500, September 2001.

\bibitem[Gymrek et~al.(2013)Gymrek, McGuire, Golan, Halperin, and
  Erlich]{Gymrek2013}
M.~Gymrek, A.~L. McGuire, D.~Golan, E.~Halperin, and Y.~Erlich.
\newblock {Identifying Personal Genomes by Surname Inference}.
\newblock \emph{Science}, 339\penalty0 (6117):\penalty0 321--324, January 2013.
\newblock ISSN 0036-8075.

\bibitem[Hall and Dowling(1980)]{Hall1980}
Patrick A.~V. Hall and Geoff~R. Dowling.
\newblock {Approximate String Matching}.
\newblock \emph{ACM Computing Surveys}, 12\penalty0 (4):\penalty0 381--402,
  December 1980.
\newblock ISSN 03600300.

\bibitem[Hamming(1950)]{Hamming1950}
Richard~Wesley Hamming.
\newblock {Error-Detecting and Error-Correcting Codes}.
\newblock \emph{Bell System Technical Journal}, 29:\penalty0 147--160, 1950.

\bibitem[Huang et~al.(2011)Huang, Evans, and Katz]{Huang2011}
Yan Huang, David Evans, and Jonathan Katz.
\newblock {Faster secure two-party computation using garbled circuits}.
\newblock \emph{USENIX Security Symposium}, 2011.

\bibitem[Huang et~al.(2012)Huang, Evans, and Katz]{Huang2012}
Yan Huang, David Evans, and Jonathan Katz.
\newblock {Private Set Intersection: Are Garbled Circuits Better than Custom
  Protocols?}
\newblock \emph{NDSS}, 2012.

\bibitem[Ingman and Gyllensten(2006)]{Ingman2006}
Max Ingman and U~Gyllensten.
\newblock {mtDB: Human Mitochondrial Genome Database, a resource for population
  genetics and medical sciences}.
\newblock \emph{Nucleic Acids Research}, 34:\penalty0 749--751, 2006.

\bibitem[Jha et~al.(2008)Jha, Kruger, and Shmatikov]{Jha2008}
Somesh Jha, Louis Kruger, and Vitaly Shmatikov.
\newblock {Towards Practical Privacy for Genomic Computation}.
\newblock In \emph{2008 IEEE Symposium on Security and Privacy (sp 2008)},
  pages 216--230. IEEE, May 2008.
\newblock ISBN 978-0-7695-3168-7.

\bibitem[Levenshtein(1966)]{Levenshtein1966}
Vladimir Levenshtein.
\newblock {Binary Codes Capable of Correcting Deletions, Insertions and
  Reversals}.
\newblock \emph{Soviet Physics Doklady}, 10:\penalty0 707, 1966.

\bibitem[Li et~al.(2007)Li, Wang, and Yang]{Li2007}
Chen Li, Bin Wang, and Xiaochun Yang.
\newblock {VGRAM: improving performance of approximate queries on string
  collections using variable-length grams}.
\newblock In \emph{Proceedings of the 33rd international conference on Very
  large data bases}, VLDB'07, pages 303--314, September 2007.
\newblock ISBN 978-1-59593-649-3.

\bibitem[Li and Homer(2010)]{Li2010}
Heng Li and Nils Homer.
\newblock {A survey of sequence alignment algorithms for next-generation
  sequencing.}
\newblock \emph{Briefings in bioinformatics}, 11\penalty0 (5):\penalty0
  473--83, September 2010.
\newblock ISSN 1477-4054.

\bibitem[Lunshof et~al.(2008)Lunshof, Chadwick, Vorhaus, and
  Church]{Lunshof2008}
Jeantine~E Lunshof, Ruth Chadwick, Daniel~B Vorhaus, and George~M Church.
\newblock {From genetic privacy to open consent.}
\newblock \emph{Nature reviews. Genetics}, 9\penalty0 (5):\penalty0 406--11,
  May 2008.
\newblock ISSN 1471-0064.

\bibitem[Naccache and Stern(1998)]{Naccache1998}
David Naccache and Jacques Stern.
\newblock {A new cryptosystem based on higher residues}.
\newblock \emph{Proceedings of the 5th ACM conference on on computer and
  communication security}, pages 59--66, 1998.

\bibitem[Paillier(1999)]{Stern1999}
Pascal Paillier.
\newblock {Public-Key Cryptosystems Based on Composite Degree Residuosity
  Classes}.
\newblock \emph{Advances in Cryptography - Eurocrypt '99}, 1592:\penalty0
  223--238, 1999.

\bibitem[Papapetrou et~al.(2010)Papapetrou, Siberski, and
  Nejdl]{Papapetrou2010}
Odysseas Papapetrou, Wolf Siberski, and Wolfgang Nejdl.
\newblock {Cardinality estimation and dynamic length adaptation for Bloom
  filters}.
\newblock \emph{Distributed and Parallel Databases}, 28\penalty0
  (2-3):\penalty0 119--156, September 2010.
\newblock ISSN 0926-8782.

\bibitem[Schnell et~al.(2009)Schnell, Bachteler, and Reiher]{Schnell2009}
Rainer Schnell, Tobias Bachteler, and J\"{o}rg Reiher.
\newblock {Privacy-preserving record linkage using Bloom filters.}
\newblock \emph{BMC medical informatics and decision making}, 9\penalty0
  (1):\penalty0 41, January 2009.
\newblock ISSN 1472-6947.

\bibitem[Smith and Waterman(1981)]{Smith1981}
Temple~F. Smith and Michael~S. Waterman.
\newblock {Identification of common molecular subsequences.}
\newblock \emph{Journal of molecular biology}, 147\penalty0 (1):\penalty0
  195--7, March 1981.
\newblock ISSN 0022-2836.

\bibitem[Troncoso-Pastoriza et~al.(2007)Troncoso-Pastoriza, Katzenbeisser, and
  Celik]{Troncoso-Pastoriza2007}
Juan~Ram\'{o}n Troncoso-Pastoriza, Stefan Katzenbeisser, and Mehmet Celik.
\newblock {Privacy preserving error resilient dna searching through oblivious
  automata}.
\newblock In \emph{Proceedings of the 14th ACM conference on Computer and
  communications security - CCS '07}, page 519, New York, New York, USA,
  October 2007. ACM Press.
\newblock ISBN 9781595937032.

\end{thebibliography}

\graphicspath{{images/}}

\end{document}